\documentclass[fleqn,12pt]{article}
\usepackage{amssymb,amsmath}
\newcommand{\slap}{\partial\hspace{-0.55em}/}
\begin{document}
\baselineskip 17pt
\begin{flushright}
KEK-TH-679
\end{flushright}
\begin{center}
{\large{\bf 
Lepton Flavor Violation in the Randall-Sundrum Model \\
with Bulk Neutrinos}} 

\vspace{1cm}

{\large Ryuichiro Kitano\footnote
{email: {\tt ryuichiro.kitano@kek.jp}}}

\vspace{.5cm}

{\small {\it Theory Group, KEK, Oho 1-1, Tsukuba, Ibaraki 305-0801,Japan \\
and \\
Department of Particle and Nuclear Physics,
The Graduate University for Advanced Studies,\\
Oho 1-1, Tsukuba, Ibaraki 305-0801, Japan}}

\vspace{1.5cm}
{\bf Abstract}
\end{center}

\bigskip
Lepton flavor violation
in the Randall-Sundrum model with
bulk neutrinos is considered.
Grossman and Neubert
recently proposed that
the existence of tiny neutrino masses and large mixing 
could be explained by the presence of the bulk neutrinos 
in the Randall-Sundrum background.
Since the zero mode wave functions of 
the bulk neutrinos are localized on Planck brane,
the Dirac Yukawa couplings on our brane are 
exponentially suppressed
enough to generate tiny neutrino masses.
However, the existence of Kaluza-Klein modes 
of these bulk neutrinos enhance 
lepton flavor violating processes such as
$\mu \to e \gamma$,
from which lower bounds on their masses can be derived.
We find that the first KK mode must be heavier than 25 TeV
if all the neutrino Yukawa couplings
are of order unity, which requires 
a fine-tuning for the Higgs mass parameter.\\
PACS Numbers: 11.10.Kk, 13.35.Bv
\newpage

Recently,
it was pointed out that
the existence of extra-dimensions could be
a solution of the hierarchy problem
\cite{Arkani-Hamed:1998rs,Randall:1999ee}.
In the Randall-Sundrum (RS) model,
the electroweak scale is generated from
the Planck scale parameters
through the non-factorizable metric of the extra-dimension.
In these model, however,
it is not easy to explain small neutrino masses
suggested by the neutrino oscillation data.
In four dimensional theory,
the seesaw mechanism is an elegant way of providing 
such tiny neutrino masses \cite{seesaw}.
However, the RS scenario is incompatible with
an intermediate scale such as the right-handed Majorana mass scale
because all the dimensionful parameters on weak brane are
suppressed by the exponential factor to 
electroweak scale.
Recently a new mechanism for generating
tiny neutrino masses was 
proposed by Grossman and Neubert
\cite{Grossman:1999ra}.
They introduce bulk fermion fields
which couple to the lepton doublets
with Dirac Yukawa couplings in the same way
as right-handed neutrinos.
In this scenario,
the neutrino Dirac Yukawa couplings are 
exponentially suppressed 
in four dimensions
because
the zero mode wave functions of the bulk fermions
are localized at the Planck brane
and the tiny Dirac mass terms for
neutrinos are generated.

The introduction of bulk fermions
in the RS background
leads to the existence of Kaluza-Klein (KK) modes whose masses are 
at the electroweak scale in four dimensional theory,
and the neutrino Dirac Yukawa
couplings for these KK modes are not suppressed.
Therefore sizable effects may arise
in the phenomenology at the electroweak scale 
\cite{Chang:1999nh,Davoudiasl:1999tf,Pomarol:1999ad}
because of the large mixing between
KK modes and neutrinos.
Grossman and Neubert pointed out 
that the invisible width of the $Z^0$ boson
restricts this mixing.

In this paper,
we consider the lepton flavor violating processes
in the RS model with bulk neutrinos.
The neutrino oscillation data implies that
lepton flavor symmetry is violated.
In the standard model with neutrinos,
lepton flavor violation in the charged lepton sector
is too small to be observed
e.g.\ Br$(\mu \to e \gamma) \sim 10^{-40}$ \cite{Petcov:1977ff}.
However with KK modes,
the experimental bound for Br$(\mu \to e \gamma)$ 
gives severe constraints on the mixing between
the bulk fermions and neutrinos.
If the Yukawa coupling is $O(1)$,
the KK modes should be heavier than 25 TeV
which means the fine-tuning of the Higgs mass parameter
is necessary to reproduce  correct vacuum expectation values (VEVs).

The RS model
is a five dimensional theory
in which the fifth dimension is compactified to $S_1/{\bf Z}_2$
and two 3-branes are located at two orbifold fixed points.
The five dimensional metric is given as
\begin{eqnarray}
ds^2 = e^{-2kr_c |\phi|} \eta_{\mu \nu} dx^\mu dx^\nu
-r_c^2 d\phi^2 \ ,
\end{eqnarray}
where $k$ is a parameter of the order of the fundamental scale $M$,
$r_c$ is the compactification radius 
which is also of the order of $M^{-1}$
and
$\phi$ is the coordinate of the fifth dimension
which is defined on $[0,\pi]$.
The standard model fields are confined 
on the brane at $\phi=\pi$.
In this set-up,
the Planck scale $M_{\rm Pl}$ and 
the VEV for the Higgs field $v$
in the four dimensional effective theory
are given as
\begin{eqnarray}
M_{\rm Pl}^2 = \frac{M^3}{k} (1-e^{-2k r_c \pi}) \ ,\ \ \ 
v=e^{-k r_c \pi} v_0 \ , \label{1}
\end{eqnarray}
where $v_0$ is the VEV
for the Higgs field in the five dimensional theory.
We can see from eq.(\ref{1}) that
if we consider $M,k,1/r_c$ and $v_0$ as parameters
of the order of $10^{19}$ GeV,
all the dimensionful parameters on weak brane
such as $v$ can be 
taken of the order of $100$ GeV
for $k r_c \sim 12$ 
while keeping $M_{\rm Pl}$ close to $10^{19}$ GeV.

The action of the bulk fermion 
in the RS set-up is given by \cite{Grossman:1999ra}
\begin{eqnarray}
S=\int d^4 x \int_0^\pi d\phi
\sqrt{G} \left \{
\frac{i}{2} \bar{\Psi} \hat{\gamma}^A \partial_A \Psi
-\frac{i}{2}(\partial_A \bar{\Psi}) \hat{\gamma}^A \Psi
-m \  {\rm sgn} (\phi) \bar{\Psi} \Psi
\right \} \ . \label{3}
\end{eqnarray}
The Majorana mass term is omitted by imposing
lepton number conservation.
Kaluza-Klein decomposition of $\Psi$ reduces this action
to the usual four dimensional Dirac action which is given by
\begin{eqnarray}
S=\sum_n \int d^4 x
\left \{ 
\bar{\psi}_n(x) i \slap \psi_n (x)
-m_n \bar{\psi}_n (x) \psi_n (x)
\right \}, \label{2}
\end{eqnarray}
where $\psi \equiv \psi^L + \psi^R$,
and $\psi^{L,R}$ is defined as
\begin{eqnarray}
\Psi_{L,R} (x,\phi) \equiv \frac{1}{2} (1 \mp \gamma_5) \Psi
= \frac{e^{2kr_c |\phi|}}{\sqrt{r_c}} \sum_n \psi^{L,R}_n (x)
\hat{f}_n^{L,R} (\phi) \ .
\end{eqnarray}
The wave functions $\hat{f}_n^{L,R}$ and 
the masses of KK modes $m_n$ are expressed
by the following parameters:
\begin{eqnarray}
\epsilon=e^{-k r_c \pi} \ ,\ t=\epsilon e^{k r_c |\phi|}
\ ,\ \nu=\frac{m}{k}\ , \label{6}
\end{eqnarray}
where the parameter $\epsilon$ should be fixed 
as $\epsilon \sim 10^{-16}$ i.e.\ $k r_c \sim 12$ 
in order to produce the hierarchy 
between the Planck scale and the electroweak scale.
The parameter $t \in [\epsilon,1]$ is 
the redefined spatial coordinate of the fifth dimension;
$t=\epsilon$ is the location of the Planck brane
and $t=1$ is that of our brane.
By rescaling the function as 
$\hat{f}_n^{L,R} (\phi) \to \sqrt{k r_c \epsilon} f_n^{L,R} (t)$,
the wave functions and the masses are explicitly given as
\begin{eqnarray}
f_0^L (t) =0 \ ,\ 
f_0^R(t)= \sqrt{\frac{1-2\nu}{1-\epsilon^{1-2\nu}}}  t^{-\nu} \ ,
\end{eqnarray}
\begin{eqnarray}
f_n^{L,R} (t) = \frac{\sqrt{2t}}{J_{\nu+\frac{1}{2}}(x_n)} 
J_{\nu \mp \frac{1}{2}} (x_n t) \ \ \ (n \neq 0\ ,\ \nu>\frac{1}{2})\ ,
\end{eqnarray}
\begin{eqnarray}
m_n= \epsilon k x_n \ , \label{neo9}
\end{eqnarray}
where $x_n$ is the solution of
$J_{\nu - \frac{1}{2}} (x_n) =0$.
For $\epsilon \sim 10^{-16}$,
the masses of the KK modes $m_n$ are of the order of
the electroweak scale.
The zero mode wave function
(there is a zero mode $f_0^R$ since $x_0=0$.)
on our brane is very suppressed such as
$f_0^R (1) \propto \epsilon^{\nu-\frac{1}{2}}$
for $\nu > 1/2$.
This smallness is the origin of the tiny neutrino masses.
The wave functions of KK modes are not suppressed 
($f_n^R (1)=\sqrt{2}$) and this gives
large coupling constants which violate lepton flavor symmetry. 

In this theory, even numbers of bulk fermions are needed
to cancel the parity anomaly of the ${\bf Z}_2$ orbifold
symmetry \cite{Redlich:1984kn,Callan:1985sa}.
For simplicity, 
we consider the minimal case in which
there are two bulk fermions $\Psi^\alpha \ (\alpha=1,2)$.

We can construct
the gauge invariant interaction terms between 
the lepton doublets 
$L_0^i = (\nu_{0L}^i,e_{0L}^i)$ $(i=e,\mu,\tau)$ 
and the bulk fermions $\Psi^\alpha \ (\alpha=1,2)$ as
\begin{eqnarray}
S_Y=-\int d^4 x e^{-4k r_c \pi}
\left \{
\hat{y}_{i\alpha} \bar{L}_0^i (x) 
\widetilde{H}_0 (x) \Psi_R^\alpha (x,\pi) + {\rm h.c.}
\right \} \ ,
\end{eqnarray}
where $\widetilde{H_0} \equiv i\sigma_2 H_0^* =(H_0^{0*},H_0^-)$ 
is Higgs field.
The couplings $\hat{y}_{i\alpha}$ are dimensionful parameters
which are naturally of the order of $M_{\rm Pl}^{-1/2}$.
In four dimensional effective theory,
this action is written in terms of $\psi$ as
\begin{eqnarray}
S_Y=-\sum_{n \ge 0} \int d^4 x
\left \{
y_n^{i\alpha} \bar{L}^i (x) \widetilde{H} (x) \psi_{n,\alpha}^{R} (x) 
+{\rm h.c.}
\right \} \ ,\label{11}
\end{eqnarray}
where the lepton doublets $L$ and 
the Higgs doublet $\widetilde{H} \equiv i \sigma_2 H^*$ are 
properly rescaled
to give canonical kinetic terms in four dimensions.
The relation of $\hat{y}_{i\alpha}$ and $y_n^{i\alpha}$
is
\begin{eqnarray}
y_n^{i\alpha}
=\sqrt{k} \hat{y}_{i\alpha} 
f_n^R (1) \equiv z_{i\alpha} f_n^R (1)\ . \label{7}
\end{eqnarray}
Here we take that $z_{i\alpha}$ to be parameters of order unity.
From eq.(\ref{2}) and eq.(\ref{11}),
the mass matrix for neutrinos is given by
\begin{eqnarray}
M=
\bordermatrix{
&\psi_{0,\alpha}^R & \psi_{1,\alpha}^R & \cdots 
& \psi_{n,\alpha}^R & \cdots \cr
\nu_L^i & v y_0^{i\alpha} & v y_1^{i\alpha} 
&\cdots & v y_n^{i\alpha} & \cdots \cr
\psi_{1,\alpha}^L & 0   & m_{1,\alpha}  &  0       &   0   & 0 \cr
\ \vdots  & 0 &  0     & \ddots &  0  &0 \cr
\psi_{n,\alpha}^L& 0 &   0    &  0      & m_{n,\alpha} &0 \cr
\ \vdots & 0&  0       &  0       &  0  & \ddots
} \ , \label{8}
\end{eqnarray}
where $m_{n,\alpha}$ (from eq.(\ref{2})) is the mass of the 
$n$-th KK mode for $\Psi^\alpha$.
For appropriate choices of $\nu$ ($\nu \sim 1$),
$y_0^{i\alpha} (\equiv z_{i\alpha} f_0^R(1))$ becomes small
enough to explain the tiny neutrino masses.
In four dimensional effective theory,
this model also contains a series of
vector-like neutrinos
which may lead to sizable lepton flavor violation 
\cite{Cheng:1977nv}.

The $3 \times 2$ submatrix for the light neutrinos is written as
\begin{eqnarray}
M_{\rm sub} = \left(
\begin{array}{ll}
 \sqrt{2\nu_1 -1} \epsilon^{\nu_1-\frac{1}{2}} z_{e1} v&
 \sqrt{2\nu_2 -1} \epsilon^{\nu_2-\frac{1}{2}} z_{e2} v \\
 \sqrt{2\nu_1 -1} \epsilon^{\nu_1-\frac{1}{2}} z_{\mu 1} v&
 \sqrt{2\nu_2 -1} \epsilon^{\nu_2-\frac{1}{2}} z_{\mu 2} v \\
 \sqrt{2\nu_1 -1} \epsilon^{\nu_1-\frac{1}{2}} z_{\tau 1} v&
 \sqrt{2\nu_2 -1} \epsilon^{\nu_2-\frac{1}{2}} z_{\tau 2} v \\
\end{array}
\right) \ ,\label{14}
\end{eqnarray}
where $\nu_\alpha=m_\alpha/k\ (\alpha=1,2)$ and
$m_\alpha$ is the mass of the bulk fermions (see eq.(\ref{3})).
We take here $m_1>m_2$.
Since this matrix is $3 \times 2$,
one of the three light neutrinos remains massless.

Now we discuss the lepton flavor violating processes 
such as 
$\mu \to e \gamma$, $\tau \to \mu \gamma$ and 
$\tau \to e \gamma$.
The experimental bounds for these processes
give severe constraints on the mass of 
the KK modes
and/or couplings $z_{i\alpha}$.

The four dimensional gauge and Yukawa
interaction terms
relevant to $\mu \to e \gamma$, $\tau \to \mu \gamma$ and
$\tau \to e \gamma$ process
are given by
\begin{eqnarray}
{\cal L}^{\rm gauge}&=&
\sum_{i=e,\mu,\tau} \frac{g}{\sqrt{2}} W_\mu^\dagger 
\bar{e}_L^i \gamma^\mu P_L \nu_L^i + {\rm h.c.} \nonumber \\
&=& 
\sum_{i=e,\mu,\tau} \sum_{A=1}^{2N+3} \frac{g}{\sqrt{2}} U_{iA}
W_\mu^\dagger \bar{e}_L^i \gamma^\mu P_L
\psi_\nu^A +{\rm h.c.} \ ,
\end{eqnarray}
\begin{eqnarray}
{\cal L}^{\rm Yukawa}
&=& \sum_{n=0}^N \sum_{i=e,\mu,\tau} \sum_{\alpha=1,2}
y_{n}^{i\alpha}
\bar{e}_L^i H^- P_R \psi_{n,\alpha}^R
-\sum_{i=e,\mu,\tau} f^i \bar{\nu}_L^i H^+ P_R e_R^i +{\rm h.c.}
\nonumber \\
&=& \sum_{n=0}^N \sum_{i=e,\mu,\tau} \sum_{\alpha=1,2}
\sum_{A=1}^N
y_{n}^{i\alpha} V_{(n,\alpha)A} \bar{e}_L^i H^- P_R \psi_\nu^A
\nonumber \\
&&-\sum_{i=e,\mu,\tau} \sum_{A=1}^N
f^i U_{iA}^* \bar{\psi}_\nu^A H^+ P_R e_R^i + {\rm h.c.}\ ,
\end{eqnarray}
where the indices $i,A,n,\alpha$ represent
the flavor, the mass eigenstates of neutrinos,
the KK excitations and the species of bulk fermions,
respectively.
Here the left-handed mixing matrix $U$ and
the right-handed mixing matrix $V$ are
defined as the matrices which diagonalize
$MM^\dagger$ and $M^\dagger M$, respectively.
To cut-off the infinite KK modes
we introduce $N$ and consider up to
$N$-th KK mode.
Then $U$ and $V$ are $(2N+3) \times (2N+3)$,
$(2N+2) \times (2N+2)$ matrices respectively.
The coupling $f^i$ is the lepton Dirac Yukawa coupling
of the $i$-th generation and 
$y_{n}^{i\alpha}$ are neutrino Dirac Yukawa couplings defined
in eq.(\ref{11}).
The field $\psi_\nu^A$ represents the $A$-th
mass eigenstate of the neutrinos. 

We first calculate Br($\mu \to e \gamma$).
Br($\tau \to \mu \gamma$) and Br($\tau \to e \gamma$)
can be calculated in the same way. 
The decay amplitude of $\mu \to e \gamma$ is
generally given by
\begin{eqnarray}
T(\mu \to e \gamma)= e \epsilon^{\alpha *}
\bar{u}_e (p-q) \left[
i \sigma_{\alpha \beta} q^\beta
(A_L P_L + A_R P_R)
\right] u_\mu (p) \ , \label{12}
\end{eqnarray}
where $P_L$ and $P_R$ are the 
chiral projection operators.
The decay width is given by
\begin{eqnarray}
\Gamma (\mu \to e \gamma)=\frac{e^2}{16 \pi}m_\mu^3
(|A_L|^2+|A_R|^2) \ ,
\end{eqnarray}
where $m_\mu$ is the mass of the muon.
Neglecting the mass of the electron,
$A_L$ and $A_R$ can be expressed as
\begin{eqnarray}
A_L=0 \ ,
\end{eqnarray}
\begin{eqnarray}
A_R=\frac{g^2}{(4\pi)^2} \sum_A
&& \hspace*{-.6cm}
\frac{m_\mu}{M_W^2} U_{eA} U^*_{\mu A} \nonumber \\
&& \hspace*{-1.8cm}
\times \frac{1}{24(1-\xi_A)}
(10-43 \xi_A +78 \xi_A^2 -49 \xi_A^3 +4 \xi_A^4
+18\xi_A^3 \log \xi_A)\ , \\
\nonumber \\
(\xi_A \equiv m_A^2/M_W^2) \nonumber 
\end{eqnarray}
where $M_W$ is $W$ boson mass
and $m_A$ is the mass of the $A$-th mass eigenstates of neutrinos.
Notice that this model predicts $\mu^- \to e^-_L \gamma$ 
(or $\mu^+ \to e_R^+ \gamma$) decay.
If all the neutrino masses are small,
this amplitude is suppressed by the GIM mechanism
\cite{Glashow:1970gm}.
However, due to the existence of heavy neutrinos, 
the GIM cancellation does not work
and $A_R$ is estimated approximately as
\begin{eqnarray}
A_R \simeq \frac{m_\mu}{(4\pi)^2}
\sum_{n=1}^N \left(
\frac{z_{e1} z_{\mu 1}}{m_{n,1}^2} 
+ \frac{z_{e2} z_{\mu 2} }{m_{n,2}^2} \right) \ . \label{20}
\end{eqnarray}
and from eq.(\ref{neo9}),
this is written as
\begin{eqnarray}
&&A_R \simeq \frac{1}{(4\pi)^2} \frac{m_\mu}{(\epsilon k)^2}
\left(
{z_{e1} z_{\mu 1}} C_1(N)
+ {z_{e2} z_{\mu 2}} C_2 (N)
\right) \ ,\label{22} \\
&&C_\alpha (N) = \sum_{n=1}^N \frac{1}{x_{n,\alpha}^2}
\ ,\ (\alpha=1,2)\ ,
\end{eqnarray}
where $x_{n,\alpha}$ are the solutions for 
$J_{\nu_\alpha -\frac{1}{2}} (x_{n,\alpha})=0$.
The functions $C_{\alpha} (N)$ are slowly increasing
functions of $N$, and therefore
the cut off dependence of Br($\mu \to e \gamma$)
is small. 
In the limit of $N \to \infty$,
$C_{\alpha} (N) \to (2(2\nu_\alpha+1))^{-1}$.

The branching ratio is given by eq.(\ref{22}) as
\begin{eqnarray}
{\rm Br}(\mu \to e \gamma) \simeq 0.0037
\left( \frac{v}{\epsilon k} \right)^4
\left| z_{e1} z_{\mu 1} C_1(N) + z_{e2} z_{\mu 2} C_2 (N)
\right|^2 \ .
\label{17}
\end{eqnarray}
The same calculation for $\tau \to \mu \gamma$ and 
$\tau \to e \gamma$ gives
\begin{eqnarray}
{\rm Br}(\tau \to \mu \gamma) \simeq 0.00065
\left( \frac{v}{\epsilon k} \right)^4
\left| z_{\mu1} z_{\tau 1} C_1(N) + z_{\mu2} z_{\tau 2} C_2 (N)
\right|^2 \ ,
\label{18}
\end{eqnarray}
\begin{eqnarray}
{\rm Br}(\tau \to e \gamma) \simeq 0.00065
\left( \frac{v}{\epsilon k} \right)^4
\left| z_{e1} z_{\tau 1} C_1 (N)+ z_{e2} z_{\tau 2} C_2 (N)
\right|^2 \ .
\label{neo19}
\end{eqnarray}
The present experimental bounds are
Br$(\mu \to e \gamma)<1.2 \times 10^{-11}$ \cite{Brooks:1999pu}, 
Br$(\tau \to \mu \gamma) < 1.1 \times 10^{-6}$ \cite{Ahmed:1999gh} and
Br$(\tau \to e \gamma) < 2.7 \times 10^{-6}$ \cite{Edwards:1997te}.
If all $z_{i\alpha}$ are of order unity,
the dimensionless combination $v/\epsilon k$ must satisfy
\begin{eqnarray}
\frac{v}{\epsilon k} \lesssim 0.02 \ , \label{28}
\end{eqnarray}
from the constraint from Br($\mu \to e \gamma$).
In eq.(\ref{28}), we use $C_\alpha (N) \sim (2(2\nu_\alpha+1))^{-1}$ and
$\nu_\alpha \sim 1$ which
is a reasonable region for producing light neutrino masses.
The parameter $x_{1,\alpha}$ is roughly estimated to be 
$x_{1,\alpha} \sim 3$ for $\nu_\alpha \sim 1$,
so that we can derive the following bound for the lowest KK mode 
$m_{\rm KK}$ from eq.(\ref{neo9}) and eq.(\ref{28}):
\begin{eqnarray}
m_{\rm KK} \gtrsim 25 \ {\rm TeV}\ .
\end{eqnarray}
Since this value is two order of magnitude larger than
the Higgs VEV, a fine-tuning of $10^{-2}$
is necessary.

An individual constraint on $z_{i\alpha}$ can be obtained 
by considering the neutrino oscillation data.
To reproduce the mixing angle 
$\sin^2 2\theta_{12}\sim 10^{-2}$ for the small angle MSW solution,
$\sin^2 2\theta_{12}\sim 1$ for the large angle MSW solution,
$\sin^2 2\theta_{23}\sim 1$ to explain the atmospheric neutrino anomaly and
$\sin^2 2\theta_{13}\lesssim 0.1$ from the CHOOZ experiment 
\cite{Fukuda:1998mi},
the structure of the Yukawa couplings are roughly given by
\begin{eqnarray}
&&|z_{e1}| \sim x |z_{\mu 1}| \sim x|z_{\tau 1}| \ ,\\
&&|z_{e2}| \ll |z_{\mu 2}| \sim |z_{\tau 2}|\ ,
\end{eqnarray}
where $x \sim 14,1/28$ for the small angle MSW solution 
and 
$x \sim 0.7$ for the large angle MSW solution \cite{Grossman:1999ra}.
Therefore the restrictions on ${z}_{i1}$ are 
given as
\begin{eqnarray}
&&  \frac{v |{z}_{e1}| }{\epsilon k} 
\lesssim 0.08\ ,\ 
\frac{v  |{z}_{\mu 1}| }{\epsilon k}
\sim \frac{v |{z}_{\tau 1}|}{\epsilon k}
\lesssim 0.006\ (x=14)
\label{27}\\
&& \frac{v |{z}_{e1}| }{\epsilon k}
\lesssim 0.004\ ,\ 
\frac{v |{z}_{\mu 1}| }{\epsilon k}
\sim 
\frac{v |{z}_{\tau 1}| }{\epsilon k} 
\lesssim 0.1\ 
(x=\frac{1}{28})\\
&& \frac{v |{z}_{e1}| }{\epsilon k}
\sim
\frac{v |{z}_{\mu 1}| }{\epsilon k}
\sim 
\frac{v |{z}_{\tau 1}| }{\epsilon k}
\lesssim 0.02\ (x=0.7)\ .
\end{eqnarray}
Since we only know the upper bound on 
the mixing angle $\sin^2 2\theta_{13}$,
typically we take $\sin^2 2\theta_{13} = 0.05$.
Then the constraints on $z_{i2}$ are given by
\begin{eqnarray}
\frac{v |{z}_{e2}|}{\epsilon k} \lesssim 0.009 \ ,\ 
\frac{v |{z}_{\mu2}|}{\epsilon k} 
\sim 
\frac{v |{z}_{\tau2}|}{\epsilon k} \lesssim 0.05\ .\label{35}
\end{eqnarray}

In ref. \cite{Grossman:1999ra},
the constraint $v |{z}_{i\alpha}|/\epsilon k \lesssim 0.1$
was derived
from the invisible decay width of the $Z^0$ boson
i.e.\ the deviation from unitarity of the MNS matrix
which is $3 \times 3$ submatrix of the matrix $U$ \cite{Maki:1962mu}.
We can find more severe constraints from considering
lepton flavor violation.
The smallness of $v |{z}_{i\alpha}|/\epsilon k$ means that the
five dimensional Yukawa couplings $\hat{y}_{i\alpha}$ 
or the five dimensional VEV of the Higgs field $v_0$
should be much smaller than $M_{\rm Pl}^{-1/2}$ or $M_{\rm Pl}$,
respectively,
which is the only natural scale of the original parameters.
In this sense,
the bounds in eq.(\ref{27}--\ref{35}) are
considered to be somewhat unnatural.

In conclusion,
we have considered lepton flavor violating processes
in the context of the small extra-dimension scenario.
The neutrino mass and mixing needs
right-handed neutrinos, but naive introduction of the 
right-handed neutrino does not provide
tiny neutrino masses.
Grossman and Neubert then proposed
the existence of right-handed neutrinos which live in bulk
and couple to the lepton doublets
and we saw that this model
lead to small Dirac neutrino mass terms.
The neutrino mass and mixing
causes lepton flavor violation.
The KK modes of the right-handed neutrinos
enhance the branching ratio of these processes.
We calculated the Br($\mu \to e \gamma$),
Br($\tau \to \mu \gamma$) and Br($\tau \to e \gamma$)
and found that Br($\mu \to e \gamma$) 
gives severe constraints on the neutrino Yukawa couplings 
$\hat{y}_{i\alpha}$
and/or the Higgs mass parameter $v_0$ in the five dimensional theory.

\section*{Acknowledgments}
The author would like to thank
Y.~Okada, J.~Hisano 
and A.~Akeroyd 
for reading manuscript 
and useful comments.
He also thanks K.~Okumura and S.~Kiyoura for discussions.

\end{document}